\documentclass[aps,prl,twocolumn,groupedaddress]{revtex4}
\usepackage{graphicx}
\usepackage{amsmath}
\usepackage{amssymb}
\usepackage{amsthm}
\usepackage{array}
\usepackage{xy}
\usepackage{subfigure}
\usepackage[pdfnewwindow]{hyperref}
\usepackage{bm}

\def\beq{\begin{equation}}
\def\eeq{\end{equation}}
\def\bea{\begin{eqnarray}}
\def\eea{\end{eqnarray}}

\pdfoutput=1

\def\cal#1{\mathcal{#1}}

\def\kt{k_{\rm B}T}

\newcommand{\location}{http://nanotheory.lbl.gov/people/design_rules_paper}

\begin{document}

\title{Design rules for the self-assembly of a protein crystal}

\author{Thomas K. Haxton}
\author{Stephen Whitelam}
\affiliation{Molecular Foundry, Lawrence Berkeley National Laboratory, Berkeley, CA 94720, USA}

\begin{abstract}
Theories of protein crystallization based on spheres that form close-packed crystals predict optimal assembly within a `slot' of second virial coefficients and enhanced assembly near the metastable liquid-vapor critical point.  However, most protein crystals are open structures stabilized by anisotropic interactions. Here, we use theory and simulation to show that assembly of one such structure is not predicted by the second virial coefficient or enhanced by the critical point.  Instead, good assembly requires that the thermodynamic driving force be on the order of the thermal energy and that interactions be made as nonspecific as possible without promoting liquid-vapor phase separation. 

\end{abstract}


\maketitle

The need to crystallize proteins for X-ray studies has spurred the development of theories of protein crystallization. These theories are largely based on the behavior of spheres with short-range isotropic attractions, a representation motivated by two observations.  First, phase diagrams for typical proteins and spherical colloids with short range attractions are structurally similar, possessing a metastable demixing transition between a vapor of solute (solute-poor solution) and a liquid of solute (solute-rich solution)~\cite{Rosenbaum1996, Rosenbaum1996b, tenWolde1997, Rosenbaum1999}.  Second, both proteins and spherical colloids tend to crystallize when the second virial coefficient, an orientationally-averaged measure of protein-protein attraction, lies in a defined `crystallization slot'~\cite{George1994, Kulkarni2003, Gibaud2011}.  On the computer, short-range isotropic spheres crystallize poorly above the metastable liquid-vapor binodal and show enhanced nucleation rates near or below it~\cite{tenWolde1997, Talanquer1998, Soga1999, Costa2002, Fortini2008, Babu2009}. Such enhancement is indeed seen in some protein solutions~\cite{Galkin2000, Pan2005, Vekilov2010twostep}. However, other experiments show disparities with this picture. Proteins can crystallize readily above the binodal~\cite{Muschol1997, Liu2010b} and experience kinetically-impaired crystallization below it~\cite{gorti2005effects}. They can also lie in the crystallization slot and not crystallize~\cite{Wilson2003}. In addition, although the structure of protein and colloid phase diagrams is similar, the microscopic nature of the stable solid is not: most proteins do not form close-packed crystals~\cite{PDB}.

These disparities motivate a theoretical approach to protein crystallization that acknowledges additional features of proteins' interactions, particularly their anisotropy~\cite{Kern2003, Doye2007, Liu2009, Whitelam2010, Romano2011, Romano2011b}.  Such studies suggest that rules for optimal assembly of open structures are different from the rules for optimal assembly of close-packed structures. Here we explicitly demonstrate this difference. We have used extensive equilibrium and nonequilibrium numerical simulations and quantitatively accurate mean-field theory to exhaustively determine the design rules for optimal assembly of a model patterned after the SbpA surface-layer protein. The latter forms a porous square lattice with a tetrameric repeat unit on the surface of the bacterium \textit{Lysinibacillus sphaericus}, and {\em in vitro} on surfaces or in solution~\cite{Sleytr2001, Sleytr2003, Sleytr2007a, Chung2010}. We impose a simple set of model protein interactions that stabilize the two condensed phases seen in experiments: {\em specific} interactions to stabilize the open crystal structure~\cite{Norville2007} and {\em nonspecific} interactions to stabilize unstructured aggregates observed on lipid bilayers~\cite{Chung2010}.  A similar distinction between orientationally specific and nonspecific interactions has been considered in models of polymer crystallization~\cite{Hu2005}.  Such a model is crucially different from isotropic models in that the same microscopic interaction does not stabilize {\em both} crystal and liquid phases. Instead, specific and nonspecific interactions independently drive distinct critical behaviors~\cite{Hu2005,hedges2011limit}. Consequently, we find that design rules for assembly differ from those of spheres. While large density fluctuations promote crystallization of close-packed spheres~\cite{tenWolde1997}, they tend to {\em inhibit} the symmetry fluctuations required to achieve assembly of the open surface-layer lattice. Further, we find that the second virial coefficient $B_2$ bounds good assembly but does not predict it.  It is intuitively reasonable that $B_2$ should bound assembly: too strong an attraction results in kinetic trapping, while too weak an attraction suppresses crystal nucleation. However, its orientational averaging renders it blind to the microscopic origin of the attraction: model proteins with identical values of $B_2$ but different combinations of specific and nonspecific interactions can assemble well, poorly, or not at all.  

Instead, we find that good assembly \emph{can} be predicted by a combination of two design rules: the thermodynamic driving force for crystallization (defined as the free energy difference between the gas and the crystal) must be $1-2 \,  \kt$, and interactions should be made as nonspecific as possible without promoting liquid-vapor phase separation. In experimental terms, our results suggest adjusting solution conditions in order to impose a defined supersaturation at the liquid-vapor binodal. While crystallization \emph{can} happen at or below the binodal, we find that such large-scale nonspecific association usually leads to slow dynamics and poor yield. Taken in the context of recent simulation work~\cite{wilber2007reversible,mike,new_rapaport}, our findings suggest that the rules governing the assembly of protein crystals are in important ways more like those governing viral capsid self-assembly than those underpinning the crystallization of spherical colloids.  

\begin{figure}[t]
\begin{center}
\includegraphics[width=\linewidth]{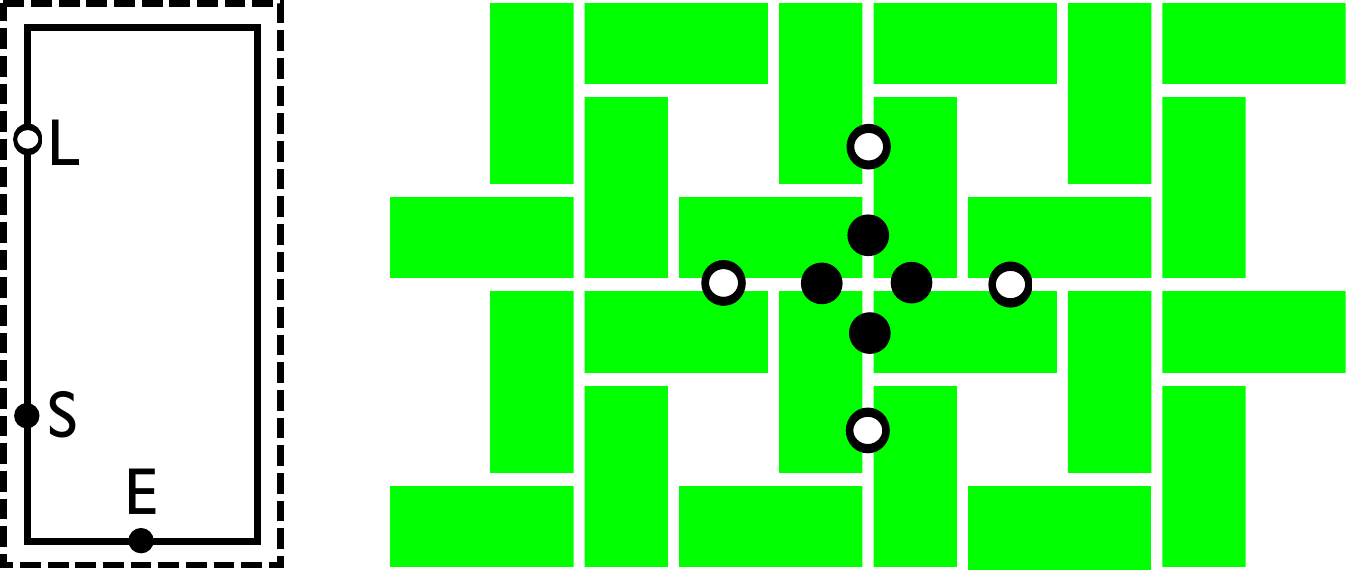}
\end{center}
\caption{(a) Monomer geometry.  (b) Square lattice stabilized by the two chemically specific interactions: internal bonds (filled circles) and external bonds (open circles).}
\label{model}
\end{figure}

{\em Model and methods.} We consider a generalization of the model SbpA surface-layer protein introduced in Ref.~\cite{Whitelam2010}. Hard rectangular monomers of width $a$ ($\equiv 3.9$ nm) and length $la$ ($l=2.2$) live on a smooth, two-dimensional substrate. Monomer interactions acknowledge the tendency of SbpA proteins to form both unstructured aggregates~\cite{Chung2010} and an open square lattice of tetramers~\cite{Norville2007}. To allow formation of the square lattice, monomers are decorated by three patches labeled E (edge), S (short arm) and L (long arm), each located on the hard-core boundary a distance $a/2$ from the nearest vertex, as shown in Fig.~\ref{model}. Patches mediate a chemically specific internal bond of energy $-\epsilon_{\rm int} \, \kt$ if the E and S patches of neighboring monomers approach closer than $\Delta=a/5$ and an external bond of energy $-\epsilon_{\rm ext} \, \kt$ if two L patches of neighboring monomers approach closer than $\Delta$.  To permit unstructured aggregation, monomers also experience a nonspecific attraction of energy $-\epsilon_{\rm n} \, \kt$ if their surrounding rectangular forcefields (of width $a+2 \Delta$ and length $la+2 \Delta$) overlap. 

To determine design rules for assembly, we extensively varied all three energetic parameters and the packing fraction $\phi$.   We will discuss the effects of separately varying $\epsilon_{\rm int}$ and $\epsilon_{\rm ext}$ elsewhere.  Here, we present results for $\epsilon_{\rm ext}/\epsilon_{\rm int}=2$ in terms of a single specific interaction parameter $\epsilon_{\rm s}\equiv\epsilon_{\rm int}=2\epsilon_{\rm ext}$.  We find that the results presented here are largely insensitive to the choice of the ratio $\epsilon_{\rm ext}/\epsilon_{\rm int}$.

We solved model thermodynamics in two ways, as detailed in the \href{\location/methods.pdf}{supplemental methods section} that is available \href{\location/methods.pdf}{\nopagebreak online}.  We used analytic mean-field theory to determine the thermodynamic driving force for assembly, phase boundaries for stable and metastable phases, and reduced second virial coefficients $B_2^{\star} \equiv B_2/B_2^{\rm hard \, core}$. $B_2 = \left( 4\pi\right)^{-1} \int d{\bm r}_{12}d\theta_{12}\,  (1-{\rm e}^{-\beta U_{12}})$ is calculated in the conventional way, integrating over the phase space of two model proteins interacting via the energy $U_{12}$.  The hard-core normalization $B_2^{\rm hard \, core}$ is obtained similarly, but for a system with no attractive interactions. We calculated phase diagrams numerically using direct coexistence and Gibbs ensemble simulations~\cite{Panagiotopoulos1989}.  We find that phase diagrams calculated by mean-field theory and simulation agree, except that simulation reveals a narrow region of thermodynamically stable liquid that the mean-field theory does not attempt to account for. 

We determined self-assembly dynamics using virtual-move Monte Carlo simulations~\cite{Whitelam2007} of 1024 monomers at constant packing fraction, starting from well-mixed conditions. Although a truly physical dynamics cannot be effected by simulations that do not explicitly represent solvent, some important aspects of real overdamped motion are retained by this algorithm: bodies move locally according to potential energy gradients, and collective diffusion constants can be scaled according to cluster size and shape. Here, we parameterized the algorithm to ensure that tightly-bound protein clusters of hydrodynamic radius $R$ diffuse according to the Stokes solution for the overdamped motion of a sphere of radius $R$, resulting in diffusive behavior for moderate to large clusters that is more realistic than that effected by basic Brownian dynamics integrators. Taking $a=3.9$ nm, $T=300$ K, and solution viscosity $\eta=1.00 \times 10^{-3}$ Pa s, each Monte Carlo (MC) cycle corresponds to 2.42 ns. 

{\em Results.} We carried out two numerical protocols, each designed to mimic a particular experiment.  First, for three selected `proteins,' each with a different balance of specific and nonspecific interactions, we determined where on the conventional temperature-concentration phase diagram yield is best. Second, we determined the microscopic mechanisms for optimal assembly by independently varying specific and nonspecific interaction strength.  Such a protocol mimics studying a large ensemble of related proteins \emph{or} varying solvent chemistry to optimize assembly for a single protein.

\begin{figure*}[t!]
\begin{center}
\includegraphics[width=\linewidth]{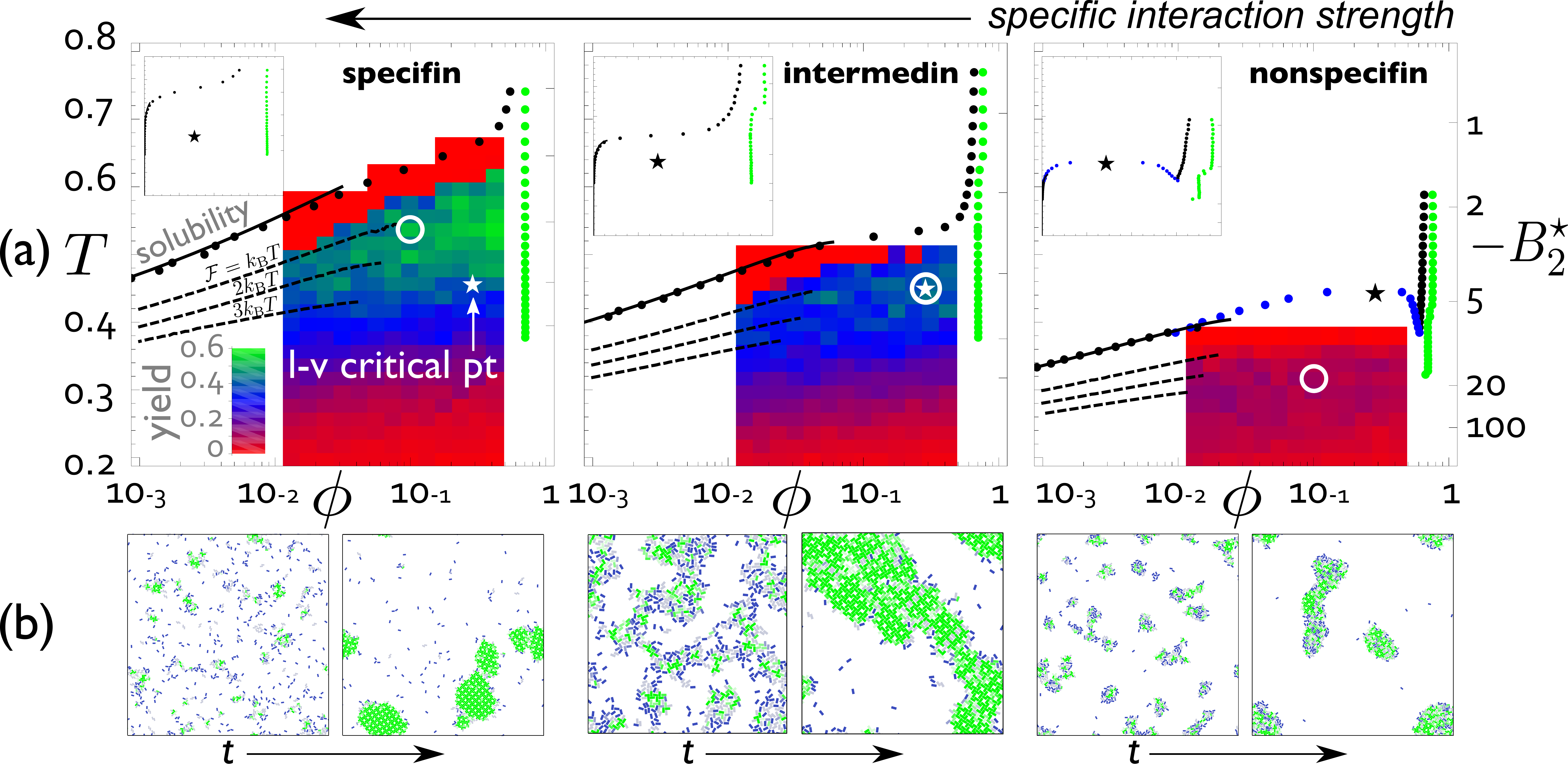}
\end{center}
\caption{{\em Optimal yield is not predicted by $B_2^{\star}$ or the position of the liquid-vapor binodal.} {\bf (a)} Phase diagrams (temperature $T\equiv 1/\epsilon_{\rm n}$ vs packing fraction $\phi$) for three model proteins whose interaction specificities decrease from left to right. We overlay the analytic solubility curve (solid), which agrees with the numeric data with no adjustable parameters, lines of fixed driving force ${\cal F}$ (dashed), and a color map of square lattice yield (obtained after dynamic simulations of $10^7$ MC cycles).  The position of best yield does not track the liquid-vapor binodal (critical points shown as stars) or a fixed slot of $B_2^{\star}$ (right ticks). Insets show phase diagrams on a more conventional linear horizontal scale.  {\bf (b)} Snapshots show example dynamic simulations for each protein after $10^5$ (left) and $5 \times 10^6$ (right) MC cycles for the points on the phase diagrams labeled by open circles.}
\label{diagrams}
\end{figure*}

In Fig.~\ref{diagrams} (a) we show temperature-concentration phase diagrams for three model proteins: {\em specifin}, with $\epsilon_{\rm s}/\epsilon_{\rm n}=2$; {\em intermedin}, with $\epsilon_{\rm s}/\epsilon_{\rm n}=1.5$; and {\em nonspecifin}, with $\epsilon_{\rm s}/\epsilon_{\rm n}=1$.  From protein to protein, the solubility curve shifts with interaction specificity more than the liquid-vapor binodal, leading to a change of phase diagram structure~\cite{Hu2005} similar to that effected by changing the range of attraction of a sphere~\cite{Gast1983,tenWolde1997,Liu2005, Pagan2005}. Specifin and intermedin display a metastable liquid-vapor coexistence, while nonspecifin displays a stable liquid-vapor coexistence. Intermedin and nonspecifin also display a transition from a square lattice ($\phi\approx 0.70$) to a close-packed crystal ($\phi\approx 0.76$) at high temperature. 

To reveal how well these proteins crystallize, we overlay the phase diagrams with color maps quantifying the crystal yield obtained after long dynamic simulations. Green indicates high yield; red, low yield. Specifin self-assembles best above the liquid-vapor critical point, intermedin assembles best near or just below it, and nonspecifin crystallizes poorly throughout its phase diagram. Below the binodal, monomers generally form kinetically sluggish gel-like or microcrystalline clusters that lead to poor yield. We illustrate dynamic trajectories leading to these outcomes in Fig.~\ref{diagrams} (b) by showing snapshots at early ($10^5$ MC cycles) and late ($5 \times 10^6$ MC cycles) times for near-optimal conditions for each protein.  (See also the corresponding online movies \href{\location/specifin.mov}{M1}, \href{\location/intermedin.mov}{M2}, and \href{\location/nonspecifin.mov}{M3}.)

For this set of proteins, optimal assembly does not track the liquid-vapor binodal.  Moreover, assembly is not predicted by the crystallization slot~\cite{George1994}, which provides a necessary but not a sufficient condition for crystallization: good assembly indeed generally occurs in a slot $-100 \lesssim B_2^{\star} \lesssim -2$, but peak yield within this slot can be highly localized (specifin), or uniformly poor (nonspecifin). From our analytic theory we calculated lines of constant ${\cal F}$, the thermodynamic driving force for crystallization. The proteins that assemble well do so in the window ${\cal F} = 1 - 2 \, \kt$.  Our analytic theory demonstrates that this window is equivalent to a supersaturation $S\equiv \phi/\phi_{\rm gas}(T)=5 - 20$, where $\phi_{\rm gas}(T)$ is the solubility packing fraction.

\begin{figure}[t!]
\begin{center}
\includegraphics[width=\columnwidth]{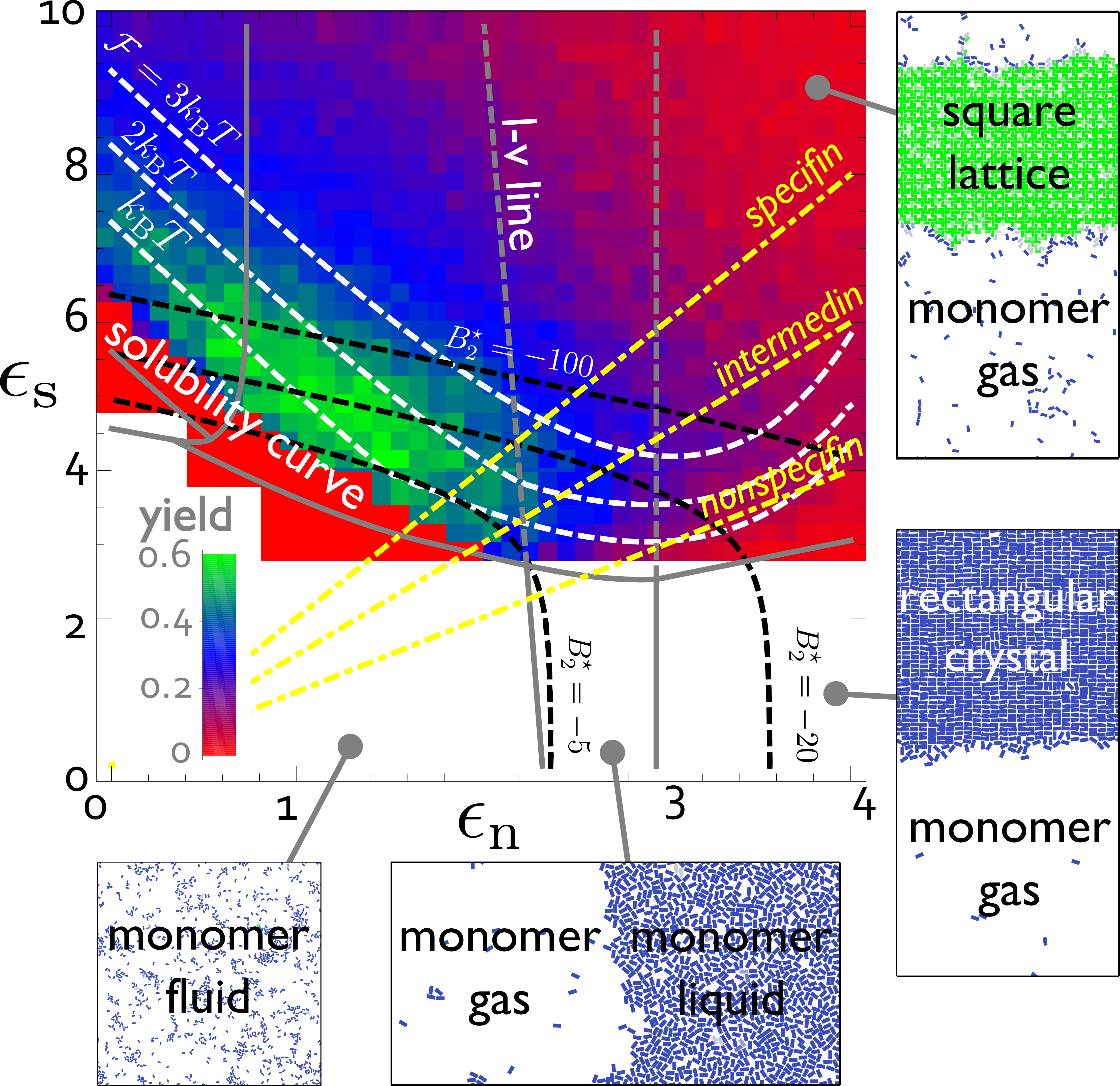}
\end{center}
\caption{{\em Best assembly occurs at moderate supersaturation and away from the liquid-vapor demixing line}.  Phase diagram and dynamic yield color map for $\phi=0.1$ and a range of protein specific and nonspecific interactions. Solid (dashed) grey curves denote the stable (metastable) boundaries for the labeled simulated coexistence combinations. All boundaries were calculated using analytic theory, except for the boundary between homogeneous and phase-separated monomer fluids; this was determined using Gibbs ensemble simulations. The dashed black (white) curves denote lines of constant $B_2^{\star}$ (driving force ${\cal F}$).  Dash-dotted yellow lines represent the proteins from Fig.~\ref{diagrams}; for a fixed protein, temperature increases to the left along these lines.}
\label{specificity1}
\end{figure}

\begin{figure}[t]
\begin{center}
\includegraphics[width=0.9\columnwidth]{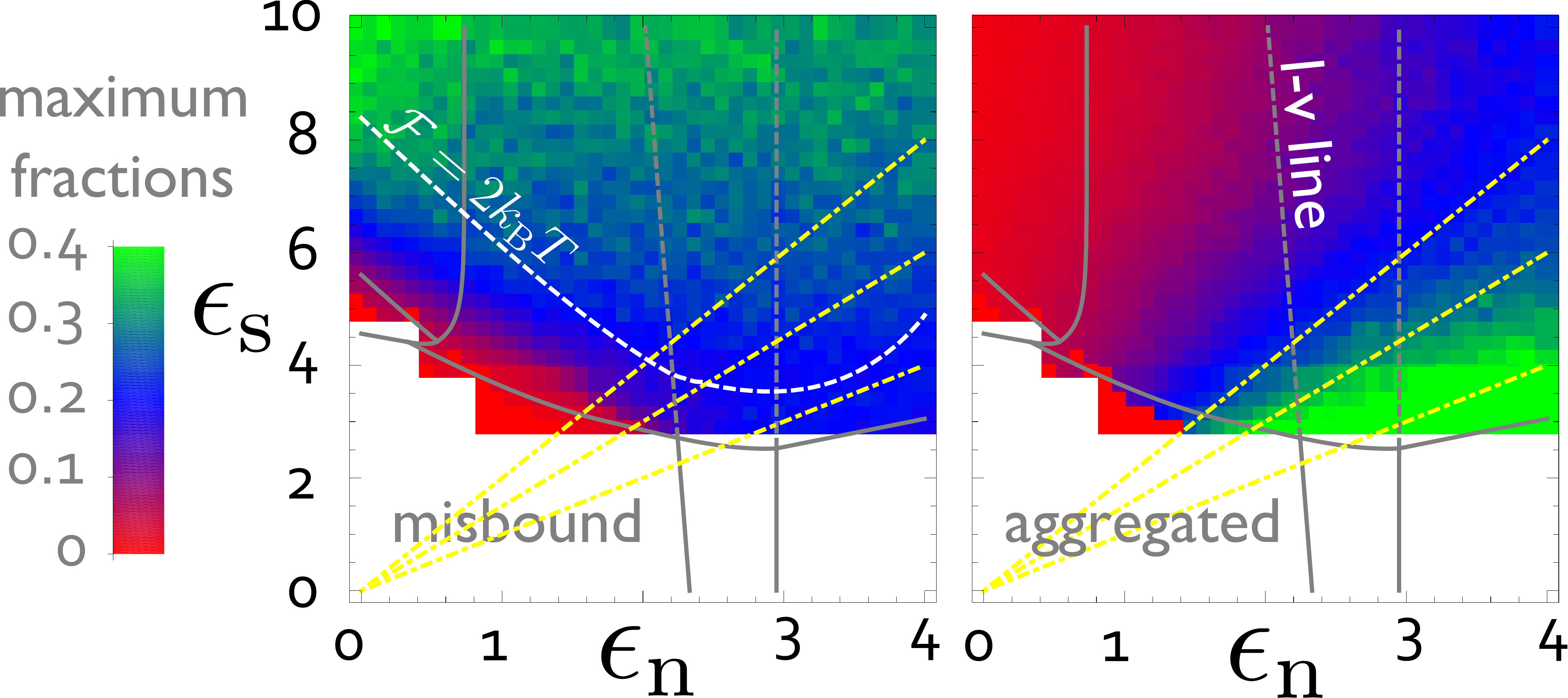}
\end{center}
\caption{{\em Large-scale nonspecific aggregation hinders crystal assembly}.  `Pathway diagrams' show color maps of the maximum fractions of misbound and nonspecifically aggregated proteins for $\phi=0.1$ and a range of protein specific and nonspecific interactions (see Fig.~\ref{specificity1}).  Comparison of the yield (Fig.~\ref{specificity1}) and pathway color maps reveals a rule of thumb for optimizing assembly: impose the strongest nonspecific interaction that does not induce liquid-vapor phase separation.}
\label{specificity2}
\end{figure}

We can clarify the molecular dynamical origins of optimal assembly by surveying an ensemble of related model proteins. We do this in Figs.~\ref{specificity1} and~\ref{specificity2} by independently varying specific and nonspecific interactions at fixed packing fraction $\phi=0.1$. Fig.~\ref{specificity1} shows a color map of crystal yield overlaid on the phase diagram spanned by the two interactions.  The surrounding simulation snapshots label the equilibrium phase or coexisting phases within each region of the phase diagram. The three proteins of Fig.~\ref{diagrams} lie on the dash-dotted yellow lines. Fig.~\ref{specificity2} shows `pathway diagrams'~\cite{hedges2011limit} which identify, along self-assembly trajectories, the maximum fractions of `misbound' proteins (those with their external specific bond satisfied but only one of their two internal specific bonds satisfied) and nonspecifically aggregated proteins (those having no specific bonds and two or more nonspecific bonds).

Optimal yield occurs in the part of the phase diagram identified by two conditions: 1) the thermodynamic driving force $\mathcal{F}$ for crystallization must be  $1 - 2 \, \kt$, and 2) the nonspecific attraction must be as large as possible without inducing liquid-vapor phase separation. The window of optimal $\mathcal{F}$ lies between the weakly supersaturated region near the solubility curve, where nucleation barriers are too high for crystallization to happen in the alloted simulation time, and the strongly supersaturated region at large $\epsilon_{\rm s}$, where misbinding predominates (see the `misbound' pathway diagram of Fig.~\ref{specificity2}). The thermodynamic driving force is substantially more predictive than the second virial coefficient; while most good assembly occurs with the displayed slot $-100 \lesssim B_2^{\star} \lesssim -5$, this slot includes large parts of the phase diagram where assembly is poor, and regions in which the target crystal is not stable.

Within the window $\mathcal{F}=1 - 2 \, \kt$, yield initially increases as specific interaction strength is traded for nonspecific interaction strength~\cite{Whitelam2010}. However, this trend terminates at the metastable liquid phase boundary. As the `aggregated' pathway diagram of Fig.~\ref{specificity2} indicates, this phase boundary signals the onset of large-scale nonspecific aggregation.  Density fluctuations associated with phase separation therefore conflict with the symmetry fluctuations required to stabilize the open crystal lattice. This behavior is strikingly distinct from that of isotropic spheres, which crystallize best at the metastable critical point. However, assembly of model capsomer proteins into icosahedral viral capsids~\cite{wilber2007reversible,mike,new_rapaport} shares the behavior of the present model; there, assembly is also impaired by liquid-like aggregation at low interaction specificity~\cite{wilber2007reversible}. We conjecture that for {\em open} structures in two and three dimensions stabilized by {\em anisotropic} attractions, weak nonspecific association should aid assembly, but large density fluctuations associated with the liquid-vapor critical point should generally impair it.

{\em Conclusions}. Typical protein phase diagrams resemble those of isotropic colloids bearing short-range attractions, but, crucially, they describe different solid structures. Here we have shown that the self-assembly of an open crystal formed by a model surface-layer protein is different in significant ways from the assembly of a close-packed crystal. However, it can be rationalized by a set of relatively simple design rules. First, the thermodynamic driving force for crystallization must be of order $\kt$. Second, interactions should be adjusted to trade specific interaction strength for {\em weak} nonspecific association; substantial nonspecific aggregation is deleterious. 

Our results suggest quantitative guidelines for optimizing crystal yield in real protein systems.  Our window of optimal thermodynamic driving force corresponds to a supersaturation of 5 to 20.  Achieving such a supersaturation without inducing large-scale nonspecific aggregation requires ensuring a large enough `metastability gap'~\cite{Asherie1996} between the solubility curve and the liquid-vapor binodal.  Inspection of the dash-dotted yellow lines in Fig.~\ref{specificity1} reveals that a protein with no metastability gap (nonspecifin) or a small metastability gap (intermedin) can be transformed into a protein with a large metastability gap (specifin) by increasing the specific interaction strength.  Such a transformation may be possible for real proteins by adjusting solvent chemistry.  For instance, recent experiments suggest that increasing multivalent salt concentration can facilitate protein crystal assembly by inducing specific contacts between proteins~\cite{Zhang2011}.

{\em Acknowledgements.} We thank Caroline Ajo-Franklin, Robert Jack, Behzad Rad, and Jeremy Schmit for discussions, and we acknowledge NERSC for computing facilities. This work was performed at the Molecular Foundry, Lawrence Berkeley National Laboratory, and was supported by the Director, Office of Science, Office of Basic Energy Sciences, of the U.S. Department of Energy under Contract No. DE-AC02--05CH11231.


\end{document}